\documentstyle[multicol,pre,aps,graphicx]{revtex} 
\newcommand{\beq}{\begin{equation}}
\newcommand{\eeq}{\end{equation}}
\newcommand{\beqa}{\begin{eqnarray}}
\newcommand{\eeqa}{\end{eqnarray}}

\newcommand{\mattwo}[1]{\tiny\left(\begin{array}{cc}#1\end{array}\right)}
\newcommand{\matthree}[1]{\tiny\left(\begin{array}{ccc}#1\end{array}\right)}
\newcommand{\eq}[1]{eq.~(\ref{#1})}
\newcommand{\Eq}[1]{Eq.~(\ref{#1})}
\newcommand{\half}{\frac{1}{2}}
\newcommand{\one}{{\bf 1}}
\begin{document}
\begin{titlepage}
\begin{flushright}
LU TP 01-16\\
\vspace{2mm}
May, 2001 \\
\vspace{2mm}
\end{flushright}
\begin{center}
\vspace{30mm}
\Large
{\bf Deterministic annealing and nonlinear assignment} \\
\normalsize
\vspace{15mm}
Henrik J\"onsson\footnote{henrik@thep.lu.se} and
Bo S\"oderberg\footnote{bo.soderberg@thep.lu.se} \\
\vspace{4mm}
Complex Systems Division \\
Dept. of Theoretical Physics \\
Lund University \\
S\"olvegatan 14 \\
S-223 62 LUND \\
\vspace{20mm}
%
{\em Submitted to Physical Review E} \\
\vspace{20mm}
\end{center}
\begin{abstract}
For combinatorial optimization problems that can be formulated as
Ising or Potts spin systems, the Mean Field (MF) approximation yields
a versatile and simple ANN heuristic, Deterministic Annealing.  \\ For
assignment problems the situation is more complex -- the natural
analog of the MF approximation lacks the simplicity present in the
Potts and Ising cases.  \\ In this article the difficulties associated
with this issue are investigated, and the options for solving them
discussed. Improvements to existing Potts-based MF-inspired heuristics
are suggested, and the possibilities for defining a proper variational
approach are scrutinized.
\end{abstract}
\vspace{1cm}
PACS number(s): 05.10.-a, 75.10.Hk

\end{titlepage}

\twocolumn
\section{Introduction}

Many mathematical methods originating from theoretical physics have
found use in completely different contexts, among them the variational
approach to the thermodynamics of complicated systems, lying at the
basis of e.g. the mean field approximation to spin systems. This has
been successfully used in heuristic methods in the context of
combinatorial optimization, for problems that allow a simple
formulation in terms of Ising or Potts spins. For other kinds of
combinatorial optimization problems, in particular assignment
problems, a similar approach is more difficult to achieve; the related
difficulties is the focus of this paper.

In an instance of a combinatorial optimization problem, a cost
function is defined in terms of a set of discrete variables, and the
object is to find an optimal state -- a particular state of the
variables that minimizes the cost function; in other words the ground
state, if the cost function is interpreted as a Hamiltonian.
In cases where the variables are of a binary nature, such a problem
thus amounts to finding the ground state of an Ising spin system (spin
glass) with a given Hamiltonian.

For a small problem instance, an exact method can be used to solve it
exactly. In addition to more problem-specific methods,
Branch-and-Bound \cite{pap98} provides a generic class of exact methods,
where an intelligent (as opposed to exhaustive) tree-search of the
phase-space is performed, disregarding parts that can be ruled out
beforehand.
Another interesting approach is Simulated Annealing \cite{kir83},
where a standard Monte-Carlo method is used to simulate the immersion
of the system in a heat bath, starting at a high temperature, which is
slowly lowered (annealing) in the course of the simulation. In the
limit of very slow annealing, this stochastic method is guaranteed to
yield the ground state as $T \to 0$ \cite{gem84}.

For a large system, however, finding the exact ground state can be a
very time-consuming task. For a large class of problems (NP-hard),
the expected time required scales worse than any polynomial in the
system size, and the quest for the exact ground state must be given
up. Instead, one has to resort to more or less dedicated heuristics,
to meet the more modest goal of finding states with as low a cost as
possible.

Problems that can be formulated in terms of Potts or Ising spins admit
a versatile heuristic method, {\em Deterministic Annealing}, based on
the iterative solution of the equations associated with the {\em mean
field} (MF) approximation of the system at hand, combined with a slow
decrease in temperature. With the MF variables interpreted as neuron
activities, the resulting dynamics at each temperature is that of a
generalized Hopfield (or connectionist) network
\cite{hop85}. Deterministic (MF) annealing has been successfully
applied to a range of problem types, see
e.g. \cite{pet98,gol96,jon00,lag00}.

The MF approximation is most conveniently derived from a variational
approach, where the proper Boltzmann distribution based on the true
Hamiltonian is approximated by a factorized distribution, constrained
to be the product of individual single-spin distributions, each of
which can be parameterized by the corresponding single spin
average. The optimal parameters of the approximating distribution
minimize an associated free energy.

Thus, the minimization of the cost function in a discrete phase space
is replaced by the minimization of an effective cost function in a
continuous parameter space, which in suitable coordinates (the spin
averages) interpolates between the discrete states of the original
phase space.  This gives an advantage as compared to a local
optimization method confined to the discrete space, due to the
possibility of taking shortcuts.

A somewhat different type of optimization problems is given by {\em
assignment} problems, where an optimal matching (assignment) between
two sets of objects is desired, as defined by a given cost function.
While certain subclasses of assignment problems, like e.g. {\em linear
assignment} where the cost is linear in the assignment matrix, can be
solved exactly in polynomial time, the generic assignment problem is a
non-polynomial one.

For nonlinear assignment problems, an obvious generalization of the
MF-based deterministic annealing approach is lacking, mainly due to
the absence of a simple and natural analog of the MF approximation.
While a linear cost appears to be the most sensible choice for a
variational Ansatz, it does not lead to the simplicity usually
associated with the MF approximation.
Nevertheless, it is possible to exploit the linear Ansatz to define
dedicated deterministic annealing schemes for non-linear assignment,
and we will investigate the difficulties and peculiarities involved in
connection with this. A major drawback with this approach, however, is
that the time required is exponential in the problem size, and so its
practical usefulness is limited.

A popular alternative, to avoid the complexity of such an approach, is
to tweak MF annealing as defined for Potts systems to make it apply to
assignment problems. We will discuss two common methods of this type,
Potts-plus-Penalty \cite{pet89} and SoftAssign \cite{ran96}, point out
their strong and weak points, and where appropriate suggest
improvements to the existing state of the art.

To illustrate the implementation on a specific problem type, and
to gauge the effect of the suggested improvements, a suitable subset
of the methods will be applied to a small testbed of simple
applications.

The article is structured as follows:
In Sec. II, the basic idea of variational methods in general is
described.
In Sec. III, MF Annealing for a Potts system is derived from a
variational MF approximation, and briefly described.
Sec. IV contains a general discussion of assignment problems, and
defines some notation.  The polynomial problem of Linear Assignment is
briefly discussed there.
In Sec. V, we discuss the definition of proper deterministic annealing
methods dedicated to assignment problems.
Sec. VI contains a discussion of existing tweaked Potts-based MF
approaches, and suggestions for improvements.
In Sec. VII we compare some of the suggested methods on a few simple
test problems.
Finally, Sec. VIII contains our conclusions.

\section{MF in General -- Variational Approach}

The MF approximation, as it is used in MF annealing for binary and
Potts systems, is most conveniently derived from a general variational
principle.

Given a complicated cost function $H(s)$ of the variables of interest,
$s$, the idea is to approximate its associated Boltzmann distribution
$\propto \exp(-H/T)$ (at a fixed artificial temperature $T$) with one
derived from a simpler cost function $H_V(s,\lambda)$ (e.g. a linear
one), with a set of free parameters $\lambda$ (the coefficients in
the linear case). The parameter values are then determined by
minimization of the associated free energy $F_V(\lambda)$,
\beq
	F_V(\lambda) = \left\langle H \right\rangle - T S 
	\equiv -T \log Z + \left\langle H - H_V \right\rangle,
\eeq
where $\left\langle\cdot\right\rangle$ stands for an expectation value in the approximating
distribution, and $Z$ denotes the corresponding partition function
$\sum_s \exp(-H_V(s)/T)$. $S$ is the associated entropy, given by
$-\sum_s p(s) \log p(s)$, with $p(s)$ the probability of state $s$,
$p(s) \equiv \exp(-H_V(s)/T)/Z$.

The variational free energy is bounded from below by the true free
energy, $F = -T \log(Z_0) = -T \log \sum_s \exp(-H(s)/T)$.  The
condition for an extremum of $F_V$ with respect to parameter
variations $\delta \lambda$ is
\beq
	\delta F_V \equiv \left\langle \delta H_V \left(H - H_V\right) \right\rangle_c = 0,
\eeq
where $\left\langle ab\right\rangle_c$ stands for the {\em connected} expectation value (or
cumulant) $\left\langle ab\right\rangle-\left\langle a\right\rangle\left\langle b\right\rangle$. Thus, for each parameter $\lambda_a$,
we must have
\beq
\label{dfdlam}
	\left\langle \frac{\partial H_V}{\partial \lambda_a} (H - H_V) \right\rangle_c = 0.
\eeq
Although (\ref{dfdlam}) may permit multiple solutions, including
saddle-points or local minima, it is commonly used in the search for
an optimal set of parameter values.

Note that since the expectation values involve the variational
Boltzmann distribution $\propto \exp\left(-H_V/T\right)$ and hence depend on
the temperature $T$, so will the optimal parameter values.

A particularly simple special case results when the variational cost
function depends {\bf linearly} on the parameters,
\beq
	H_V(s;\lambda) \equiv \sum_a \lambda_a E_a(s).
\eeq
Then, \eq{dfdlam} for an extremum takes the simple form
\beq
	\sum_b \left\langle E_a E_b \right\rangle_c \lambda_b = \left\langle E_a H \right\rangle_c.
\eeq
This has the form of a matrix equation, $\left\langle EE\right\rangle_c\lambda=\left\langle EH\right\rangle_c$,
and a straightforward strategy for finding a solution is given by iteratively
updating the parameters according to
\beq
\label{geninv}
	\lambda \to \left\langle E E \right\rangle_c^{-1} \left\langle E H \right\rangle_c,
\eeq
followed by the corresponding updates of the expectation values
$\left\langle EE\right\rangle_c,\left\langle EH\right\rangle_c$, which depend on the parameters via the Boltzmann
distribution.

\section{MF Annealing for Potts systems}

In order to understand the problems associated with defining a
deterministic annealing approach to assignment-based problems, it is
instructive to first review how simpler types of systems are treated.

A simple $q$-state multiple-choice variable (Potts spin) is
conveniently represented by a $q$-dimensional vector ${\bf s}$, with
the allowed states represented by the $q$ principal vectors
$(1,0,0,\ldots),\;(0,1,0,\ldots),$ etc., with the position of the
single nonvanishing component indicating which state is ``on''.  These
vectors are linearly independent, and point to the corners of a
regular $q$-simplex.

As a result, any single-spin cost function can be written as a linear
function in ${\bf s}$, $H({\bf s}) = {\bf C} \cdot {\bf s}$, where
$C_a$ is the cost associated with state $a$. For a system of $N$ Potts
spins, it follows that an arbitrary cost function can be written in a
{\em multilinear} form,
\beq
	H(s) = a + \sum_{i,a} b_{ia} s_{ia}
	+ \frac{1}{2} \sum_{i,a} \sum_{j\ne i,b} c_{ia,jb} s_{ia} s_{jb}
	+ \ldots,
\eeq
where $s_{ia}$ denotes the $a$-th component of the $i$-th spin.
Solving the associated optimization problem corresponds to finding the
ground state of the system, i.e. the combination of states of the $N$
Potts spins that minimizes $H(s)$.

The MF approximation to such a system results from a variational
approach, corresponding to an optimal approximation of the non-linear
cost $H(s)$ in terms of a {\em linear} one,
\beq
	H_V(s) = \sum_i {\bf C}_i \cdot {\bf s}_i = \sum_{ia} C_{ia} s_{ia}.
\eeq
The coefficients $C_{ia}$ constitute the parameters, and are to be
chosen so as to minimize the variational free energy $F_V$.  It is
convenient to express $F_V$ in terms of the spin averages in the
variational distribution, the ({\em mean field spins}) ${\mathbf
v}_i$, with components in $[0,1]$ amounting to
\beq
\label{mfpotts1}
	v_{ia}(C) \equiv \left\langle s_{ia} \right\rangle_V =
	\frac{\exp\left(-C_{ia}/T\right)}{\sum_b \exp\left(-C_{ib}/T\right)}.
\eeq
In the MF approximation, $v_{ia}$ corresponds to the probability for
spin $i$ to be in state $a$, consistently with the identity $\sum_a
v_{ia} = 1$. The MF spins thus interpolate between the discrete states
of the original spins; in terms of them the variational free energy
evaluates to
\beq
\label{F_potts_v}
	F_V(v) = T \sum_{ia} v_{ia} \log v_{ia} + H(v),
\eeq
which can be minimized with respect to the normalized MF spins by
adding a Lagrange parameter $\lambda_i$ for the normalization of each
MF spin ${\bf v}_i$. The condition for a extremum, equivalent to
\eq{dfdlam} amounts to
\beq
	dH(v)/dv_{ia} + T (1 + \log v_{ia}) = \lambda_i,
\eeq
which, together with the normalization that fixes the $\lambda_i$
values, gives the variational coefficients $C$ up to an unimportant
constant in terms of $v$ as
\beq
\label{mfpotts2}
	 C_{ia}(v) = \frac{\partial H(v)}{\partial v_{ia}}.
\eeq
Eqs. (\ref{mfpotts1}) and (\ref{mfpotts2}) define the {\em MF
equations}.

The MF approximation corresponds to neglecting the correlations
between the different spins, since the linear variational Ansatz used
is the most general factorized distribution $P_V(s) = \prod_i
p_i(s_i)$, where the different Potts spins obey independent
distributions.

{\em MF annealing} corresponds to solving the MF equations
iteratively, starting with a high $T$, where a fixed point with
$v_{ia} \approx 1/N$ will dominate, and slowly lowering $T$. At low
enough $T$, the MF spins will be forced {\em on shell}, i.e. for
$v_{ia} \approx s_{ia} \in 0,1$, and a suggested solution can be
extracted.

\section{Assignment Problems -- General Discussion}

\subsection{Notation}

When it comes to permutation/assignment problems, we have to
distinguish between {\em single assignment} problems and {\em multiple
assignment} problems, the latter being based on several assignments.

To begin with, we will consider the simpler case of a single
assignment, where an optimal matching between two sets of $N$ objects
is desired, i.e. for each object $i$ in the first set, an object $a$
in the second set is to be chosen, such that different $i$ are
assigned to different $a$.  There are obviously $N!$ ways to
accomplish this.

A compact way to describe an assignment is by means of the associated
{\em assignment matrix}, i.e. an $N\times N$ doubly stochastic matrix
$s$ with elements in $\{0,1\}$, such that $s_{ia} = 1$ if $i$ is
assigned to $a$, and 0 otherwise (for a somewhat different encoding of
assignment problems as used in deterministic annealing, see
\cite{lag00}). Obviously, we must have precisely one unit element in
each row as well as in each column of $s$, consistently with
\beq
	\sum_a s_{ia} = 1,\; \sum_i s_{ia} = 1.
\eeq
Then, a given single assignment problem can be described in terms of a
cost function $H(s)$, which is to be minimized.

Note, however, that, in contrast to e.g. the Potts case, the most
general cost function for single assignment (for $N>2$) is {\bf not}
linear in $s$ (see Appendix); in the worst case a polynomial of degree
$N-1$ is needed.

Alternatively, the cost function can be viewed as an explicit function
over the permutation group $P_N$: Each group element $g$ is associated
with an individual cost $C_g$, defining an element of an
$N!$-dimensional {\em cost vector} $\vec{C}$.  The relation to the
formulation in terms of the assignment matrices $s$ is $C_g = H(s_g)$,
where $s_g$ is the particular assignment matrix representing $g$.

\subsection{Thermodynamics for a single nonlinear assignment}

The thermodynamics af a system consisting of a single $N\times N$
assignment with an arbitrary cost function is not difficult to
express, when formulated in terms of a cost vector $\vec{C}$ over
group space.  The assignment then acts as a single $N!$-state Potts
spin, that can be described by an $N!$-dimensional vector $\vec{S}$
with precisely one unit component, while the rest is zero: $S_g \in
\{0,1\}$, $\sum_g S_g = 1$.

At an artificial temperature $T$, the probability of a particular
state $g$ amounts to
\beq
	V_g = \left\langle S_g\right\rangle = \frac{\exp\left(-C_g/T\right)}{\sum_h \exp\left(-C_h/T\right)}.
\eeq
As $T\to 0$, the distribution gets increasingly concentrated at the
state (group element) with the lowest cost.

When viewed this way, the difficulty lies entirely in the huge number
of states involved if $N$ is large. In order to compute one component
of $\vec{V}$, one has to know the costs for all the $N!$ states. For a
generic cost function, the associated computational complexity is
non-polynomial in the size $N$ of the system.

Thus, for a generic large assignment problem, one has to make do with
some kind of heuristic.

\subsection{Linear assignment}
\label{seclinass}

Certain classes of assignment problems can be solved exactly in
polynomial time. One such class is linear assignment, where the
cost function is constrained to be linear in the assignment matrix
$s$,
\beq
\label{Hlinass}
	H(s) = \sum_{ij} c_{ij} s_{ij},
\eeq
defining an $N\times N$ {\em cost matrix} $c$.

This problem corresponds essentially to a linear programming one, and
can be solved in polynomial time, using e.g. the so called {\em
Hungarian algorithm} \cite{kuh55,pap98}, based on the fact that the
addition of terms to $c$ depending on row or column alone, $c_{ij} \to
c_{ij} + a_i + b_j$, is equivalent to adding a constant to the cost
function, $H(s) \to H(s) + \sum_i a_i + \sum_j b_j$.  This is used to
iteratively modify the cost matrix until it takes a form where it has
zeros on a set of elements corresponding to the optimal assignment,
and non-negative values elsewhere.

Unfortunately, this doesn't help when computing thermal averages at a
finite $T$; this is still a non-polynomial task. Thus, e.g., the
expectation value of $s_{ij}$ is given in terms of a matrix $M$,
obtained from $c$ by elementwise exponentiation,
\beq
\label{M}
	M_{ij} = \exp(-c_{ij}/T),
\eeq
as
\beq
\label{v_perm}
	v_{ij} \equiv \left\langle s_{ij} \right\rangle = \frac{M_{ij} P_{ij}(M)}{P(M)}.
\eeq
Here, $P(M)$ is the {\em permanent} \cite{min78} of $M$; it has some similarities
to the determinant, being the sum of all the possible products of $N$
elements in $M$, one in each row and one in each column, but with {\em
no minus signs} in contrast to the case for the
determinant. Similarly, $P_{ij}(M)$ is the {\em subpermanent} of $M$,
obtained by removing row $i$ and column $j$ from $M$, and computing
the permanent of the remaining $(N-1)\times(N-1)$ matrix.

The expression for $v_{ij}$ in \eq{v_perm} can be derived from
\beq
	\left\langle s_{ij}\right\rangle = -\frac{T}{Z} \frac{\partial Z}{\partial c_{ij}},
\eeq
where $Z$ is the partition function,
\beqa
\nonumber
	Z
	&=& \sum_g \exp\left(-\sum_{ij} c_{ij} s_{ij}(g)/T\right)
\\
\nonumber
	&=& \sum_g \exp\left(-\sum_{ij\in g} c_{ij}/T\right)
\\
	&=& \sum_g \prod_{ij\in g} \exp\left(-c_{ij}/T\right)
	= \sum_g \prod_{ij\in g} M_{ij}
	= P(M),
\eeqa
where the restriction $ij\in g$ in the sum over $ij$ means that row
$i$ and column $j$ are matched in $s(g)$ (so $s_{ij}(g)=1$). The derivative
of $Z=P(M)$ with respect to $M_{ij}$ yields $P_{ij}(M)$, which
completes the proof of \eq{v_perm}.

The combination $M_{ij} P_{ij}(M)$, appearing in \eq{v_perm}, gives
the sum of those terms in the permanent that contain the element
$M_{ij}$. Summing this over $i$ or $j$ yields $P(M)$, ensuring that
eq. (\ref{v_perm}) yields a doubly stochastic matrix $v$.

Similarly, the expectation value of the product of two elements of $s$
becomes
\beq
\label{vv_perm}
	\left\langle s_{ij} s_{kl} \right\rangle
	= \delta_{ik} \delta_{jl} \frac{M_{ij} P_{ij}(M)}{P(M)}
	+ \frac{M_{ij} M_{kl} P_{ik,jl}(M)}{P(M)},
\eeq
where $P_{ik,jl}(M)$ is a subpermanent obtained as the permanent of
the submatrix where rows $i,k$ and columns $j,l$ are removed, if $i\ne
k$ and $j \ne l$; else it is zero. Thus $M_{ij} M_{kl} P_{ik,jl}(M)$
sums up the terms in $P(M)$ that contain $M_{ij} M_{kl}$.

As an aside, replacing the permanents by determinants in the
expression (\ref{v_perm}) for $v$ would lead to the combination
$D_{ij}(M)/D(M)$, exactly corresponding to the $j,i$ element of the
matrix inverse of $M$. The elementwise product with $M$ would yield a
doubly (quasi-)stochastic $v$, where row and column sums are equal to
one, albeit with elements of both signs.

However, while the determinant of a matrix can be calculated in
polynomial time, the permanent in general can not, in spite of their
similarity -- the computational time required to compute a generic
$N\times N$ permanent is exponential in $N$ (roughly $\propto 2^N$
using e.g. Ryser's method \cite{rys63,min78}).
%

\section{Proper Variational Method for a Single Nonlinear Assignment}

For a large generic single assignment problem, an exact solution is
out of reach, and one has to make do with heuristic methods. One
possibility then is to consider a deterministic annealing approach
based on approximating the true cost function by a variational one
that is simpler.

\subsection{Linear Ansatz for $H_V$}

The most natural Ansatz for the variational cost function $H_V$ is a
linear one,
\beq
	H_V(s) = \sum_{ij} c_{ij} s_{ij},
\eeq
with the coefficients $c_{ij}$ as free parameters.

The equation (\ref{dfdlam}) for a minimum of the variational free
energy then yields:
\beq
\label{s_var_c}
	\sum_{kl} \left\langle s_{ij} s_{kl}\right\rangle_c c_{kl} = \left\langle s_{ij} H \right\rangle_c.
\eeq
In analogy with \eq{geninv}, this is a matrix equation, from which $c$
can be formally extracted as
\beq
\label{s_var_inv}
	c = \left\langle ss\right\rangle_c^{-1} \left\langle sH\right\rangle_c.
\eeq
Note that the $N^2\times N^2$ matrix $\left\langle ss\right\rangle_c$ is not fully
invertible; it always has $2N-1$ zero-modes corresponding to the
addition of redundant terms to $c$ depending on row or column index
alone. These merely yield row and column factors in the exponentiated
matrix $M$, and are of no importance for expectation values.

If $H$ is a low-order polynomial in $s$, a solution to
eq. (\ref{s_var_c}) can in principle be computed iteratively in analogy
to the iterative solution of the Potts MF eqs. (\ref{mfpotts1},
\ref{mfpotts2}), by repeating the two steps
\begin{enumerate}
\item Calculate the expectation values appearing in eq. (\ref{s_var_c});
 they depend on the present $c$ via $M$ and its permanent (and
 subpermanents), where $M_{ij} = \exp\left(-\beta c_{ij}\right)$, in analogy to
 eqs. (\ref{v_perm},\ref{vv_perm}).
\item Obtain an updated cost matrix $c$ by means of \eq{s_var_inv},
 suitably regularized with respect to the zero-modes of $\left\langle ss\right\rangle_c$.
\footnote{In cases of instability, the change in $c$ can be decreased by e.g. a factor of 1/2.}
\end{enumerate}
This can be turned into an annealing approach, by starting with a high
$T$ (low $\beta$), and decreasing $T$ slightly after every step, until
the ``MF variables'' $v_{ij} \equiv \left\langle s_{ij}\right\rangle$ have stabilized
sufficiently close to zero or one.

A major drawback of this approach is that it is only feasible if $N$
is not too large, since the computation of expectation values involves
the computation of permanents, which requires a time exponential in
$N$.

\subsection{Quadratic $H$}

The simplest non-linear function of $s$ is quadratic, so assume $H$ to
be a given quadratic plus linear function in $s$,
\beq
\label{Hquad}
	H(s) = \half \sum_{ijkl} A_{ijkl} s_{ij} s_{kl} + \sum_{ij} B_{ij} s_{ij},
\eeq
involving a symmetric tensor $A$, $A_{ijkl}=A_{klij}$, which can be
assumed to vanish for $i=k$ or $j=l$.

The variational \eq{s_var_c} corresponding to a linear $H_V(s)$ becomes
\beqa
\nonumber
	\sum_{kl} \left\langle s_{ij} s_{kl}\right\rangle_c c_{kl}
	= \half \sum_{klmn} \left\langle s_{ij} \left( s_{kl} s_{mn} \right) \right\rangle_c A_{klmn}
\\
	+ \sum_{kl} \left\langle s_{ij} s_{kl} \right\rangle_c
	\left( \sum_{mn} A_{klmn} \left\langle s_{mn} \right\rangle + B_{kl} \right).
\label{ssc_quad}
\eeqa
Denoting the first term on the RHS of \eq{ssc_quad} by $F_{ij}$, the
effective cost matrix $c$ can be formally extracted as $c = B + A
\left\langle s\right\rangle + \left\langle ss\right\rangle_c^{-1} F$, using a suitably regularized matrix inverse.

\subsection{Group theoretical aspects}

It is instructive to view an assignment problem from a
group-theoretical point of view, where the relevant group of course is
the permutation group of $N$ elements, denoted by $P_N$.

Like any functions over group space, $H$ and $H_V$ can be expressed in
a unique way as linear combinations of the matrix elements of the
irreducible representation matrices of $P_N$ (see Appendix for
details).

Requiring $H_V$ to be linear in $s$ means that its expansion is
constrained to contain only elements from the trivial and the
fundamental irreducible representations, ${\bf e}$ and ${\bf f}$; thus, it can
be written in a unique way in the form
\beq
	H_V(g) = A + \sum_{ij} B_{ij} u^f_{ij}(g),
\eeq
where $f$ stands for the fundamental representation.  This leads to
the probability $V_g \propto \exp(-H_V(g)/T)$, such that $\sum_g V_g =
1$, for a particular group element $g$. The corresponding version of
the variational \eq{dfdlam} becomes
\beqa
\label{GTvar1}
	&& A + \sum_{kl} B_{kl} \sum_g V_g u^f_{kl}(g) = \sum_g V_g H(g),
\\
\nonumber
	&& A \sum_g V_g u^f_{ij}(g) + \sum_{kl} B_{kl}
	\sum_g V_g u^f_{ij}(g) u^f_{kl}(g)
\\
	&& \; = \sum_g V_g u^f_{ij}(g) H(g),
\label{GTvar2}
\eeqa
corresponding exactly to respectively the trivial and the nontrivial
parts of \eq{s_var_c}. The trivial part $A$ can be eliminated from
(\ref{GTvar1}) and inserted into (\ref{GTvar2}), yielding
\beqa
\nonumber
	&& \sum_{kl} \left(
	\sum_g V_g u^f_{ij}(g) u^f_{kl}(g)
	- \sum_g V_g u^f_{ij}(g) \sum_h V_h u^f_{kl}(h)
	\right) B_{kl}
\\
	&& \; = \sum_g V_g u^f_{ij}(g) H(g)
	- \sum_g V_g u^f_{ij}(g) \sum_h V_h H(h),
\eeqa
which is nothing but a disguised version of \eq{s_var_c}; the sums
over $g$ with $V_g$ as a weight correspond to averages.

For the special case (\ref{Hquad}) of a quadratic $H(s)$, the
corresponding $H(g)$ is constrained to include elements from ${\bf e}$,
${\bf f}$ and two additional representations, ${\bf a}$ and ${\bf s}$ (see Appendix),
in its irrep expansion, with dimensions $d_a = (N-1)(N-2)/2$ (if
$N>2$), and $d_s = N(N-3)/2$ (if $N>3$), respectively.

Although the above analysis illuminates the linear variational
approach from a group-theoretical point of view, the resulting
formulation is not of an immediate practical interest for large $N$,
since it (at least formally) requires the complete enumeration of the
$N!$-dimensional group space.

\subsection{Proper variational methods for multiple nonlinear assignment}

Here we will briefly discuss the possibilities for treating systems of
several distinct assignments in a variational approach.

\subsubsection{General additive Ansatz}
In the case of a generic cost function of several assignments, the
most natural choice is to consider a generic {\em additive} Ansatz for
a variational cost function, corresponding to a factorized Boltzmann
distribution. In principle, this corresponds to the MF approximation
to a system of $N!$-state Potts variables, and can be treated as such.
This can be useful, even for large systems (many distinct
assignments), as long as the individual assignments are of low
dimensionalities.

\subsubsection{Linear Ansatz}
A further simplification results from requiring that the variational
cost function not only be additive in the different assignments, but
also that the contribution from each assignment be linear.

A possible strategy then is to update the cost matrix for one
assignment at a time according to \eq{s_var_inv}, considering the
single-assignment averages associated with other assignments as
constant, and recomputing the single-assignment averages associated
with the updated cost matrix before moving on to update the next
assignment.

\subsubsection{Multilinear cost, linear Ansatz}
A particularly simple special case of the above is when the exact cost
function is a {\em multilinear} function of the assignments matrices
$s$.  For the case of two assignments, $s^{(1)}$ and $s^{(2)}$, this
means a cost function $H$ of the form
\beqa
\nonumber
	&& H(s^{(1)}, s^{(2)})
	= \sum_{ij} \sum_{kl} A_{ij,kl} s^{(1)}_{ij} s^{(2)}_{kl}
\\
	&& \; + \sum_{ij} B^{(1)}_{ij} s^{(1)}_{ij}
	+ \sum_{ij} B^{(2)}_{ij} s^{(2)}_{ij}.
\eeqa
Then an additive variational cost function automatically will be
linear
\beq
	H_V(s^{(1)}, s^{(2)})
	= \sum_{ij} C^{(1)}_{ij} s^{(1)}_{ij}
	+ \sum_{ij} C^{(2)}_{ij} s^{(2)}_{ij}.
\eeq
The resulting updates then amount simply to
\beqa
\nonumber
	C^{(1)}_{ij} &=& \sum_{kl} A_{ij,kl} \left\langle s^{(2)}_{kl}\right\rangle + B^{(1)}_{ij},
\\
\\
\nonumber
	C^{(2)}_{kl} &=& \sum_{ij} A_{ij,kl} \left\langle s^{(1)}_{ij}\right\rangle + B^{(2)}_{kl},
\eeqa
where variational expectation values are understood, computable in
terms of permanents and sub-permanents of the respective cost
matrices.

The generalization to several assignments is straightforward.

\section{Potts-based MF Heuristics for Nonlinear Assignment}

Although a proper variational approach as described above appears to
be the natural choice for constructing a deterministic annealing
approach for assignment problems, a major problem is the computational
complexity involved in the required computation of permanents and
subpermanents. Indeed, for certain problem classes an instance can be
solved exactly in the time it takes to compute a single permanent.
This implies that these methods have a rather limited applicability.

Instead, alternative methods based on Potts spins have been used to
construct faster deterministic annealing methods for non-linear
assignment problems.

\subsection{Potts plus Penalty}

One such method is based on viewing the assignment matrix $s$ as an
array of $N$-state Potts spins, one for each row. Then the row
condition, $\sum_a s_{ia}=1$, is automatically fulfilled. One then
adds to $H$ a penalty term for breaking of the column normalization,
and treats the result using Potts MF annealing, based on the modified
cost function
\beq
	H(s) + \frac{\alpha}{2} \sum_a \left(1 - \sum_i s_{ia}\right)^2,
\eeq
with the penalty strength $\alpha$ suitable adjusted. (Of course, one
might equally well consider the columns as Potts spins and add
penalties for the rows.)

This approach, in what follows referred to as {\bf PPP} for Potts plus
Penalty, has been successfully applied to a number of different
problems \cite{pet98}. The problem with a soft penalty is that it
involves a delicate tuning of the coefficient $\alpha$; too small, and
improper assignments result, where two rows are mapped to the same
column; too large, and the cost is too dominated by the penalty term,
with a consequent deterioration in performance.

In PPP, the MF spins are preferably updated in a serial manner, one
row at a time.  This leads to the most stable MF dynamics, provided
$H$ is put in multilinear form. It is easy to see then that the Potts
free energy (\ref{F_potts_v}) becomes a Lyapunov function of the
dynamics at a fixed temperature. This ensures that the MF dynamics is
well-behaved in the high-$T$ domain, with the trivial high-$T$ fixed
point losing stability in a controlled manner. It also guarantees that
PPP turns into a form of local optimization in the low-$T$ limit,
ensuring the stability of an optimal assignment.

\subsection{SoftAssign}
\label{SoftAssignSec}
An obviously ugly feature of the PPP approach is the asymmetry in the
treatment of rows and columns of $s$.  This can be cured in a slightly
more advanced Potts-inspired method, the {\em SoftAssign} (or ``Double
Potts'') approach \cite{ran96}, which can be derived as a somewhat
{\em ad hoc} improvement to PPP as follows.

In PPP, the resulting MF average is given by $v_{ij} = a_i M_{ij}
b_j$, where $M_{ij}$ is given by $\exp(-(\partial H/\partial v_{ij})/T)$
and the column factor $b_j$ comes from the penalty term, $b_j =
\exp(\alpha (1-\sum_i v_{ij})/T)$. In contrast, the row factor is the
usual automatic Potts normalization factor, ensuring the exact
normalization of rows.

The idea in SoftAssign is to skip the penalty, and freely choose
positive row and column factors so as to force the exact normalization
of both rows and columns. This leads to the following problem: Given a
matrix $M$ with non-negative elements, find vectors of row and column
factors, $a$ and $b$, such that the result,
\beq
\label{v_aMb}
	v_{ij} \equiv a_i M_{ij} b_j,
\eeq
is a doubly stochastic matrix.

This in fact determines $v$ uniquely, which can be seen by defining
$x_i = \log a_i$, $y_j=\log b_j$ ($a_i,b_j$ are assumed positive), and
noting that the correct $x,y$ minimize the strictly convex function
$f(x,y) \equiv \sum_{ij} e^{x_i} M_{ij} e^{y_j} - \sum_i \left(x_i +
y_i\right)$.

However, the proper row and column factors can not be obtained as
simple, closed expressions in the matrix elements of $M$. Instead, the
desired doubly stochastic matrix $v$ is usually obtained by
iteratively modifying $M$, alternatingly normalizing rows and columns
until the resulting matrix is sufficiently close to being correctly
normalized:
\beqa
\nonumber
	{\bf i)} &\;& M_{ij} \to \frac{M_{ij}}{\sum_k M_{ik}},
\\
\nonumber
	{\bf ii)} &\;& M_{ij} \to \frac{M_{ij}}{\sum_k M_{kj}},
\\
\nonumber
	{\bf iii)} &\;& \mbox{Go to {\bf i)}}.
\label{sinkhorn}
\eeqa
This procedure, which is due to Sinkhorn \cite{sin64}, is guaranteed
to yield convergence to a unique doubly stochastic matrix $v$.

For a nonlinear problem, we can obviously identify the derivatives
$\partial H/\partial v_{ij}$ with the elements of an estimated {\em
effective cost matrix}, obtained by linearizing the cost-landscape in
the vicinity of the present point.
Note that, for a {\em linear} assignment problem, SoftAssign leads to
exactly the same initial $M$ as in \eq{v_perm}; but that a different
doubly stochastic matrix $v$ is derived from it in \eq{v_aMb}.

As an aside, the SoftAssign approach can formally be associated with
the minimization of an entity reminiscent of a free energy,
\beq
\label{F_softass}
	F(v) = T \sum_{ij} v_{ij} \log v_{ij} + H(v),
\eeq
with $v$ constrained to be doubly stochastic, e.g. by means of adding
suitable Lagrange terms (with Lagrange multipliers associated with the
row and column factors). Although $F(v)$ is superficially highly
analogous to the Potts free energy in \eq{F_potts_v}, SoftAssign does
{\em not} correspond to a proper variational approch to approximating
the true $H(s)$, mainly because the first term is not $-T$ times the
proper entropy. Nor does $v$ correspond to the expectation value of
$s$ in an approximating Boltzmann distribution.

Note that for SoftAssign (unlike the case with a proper variational
approach), the resulting dynamics is sensitive to the precise
formulation of $H(v)$ as an extrapolation of $H(s)$ to continuous
arguments $v$.

Nevertheless, SoftAssign seems theoretically more appealing than PPP,
in treating rows and columns in a symmetric manner, and in
guaranteeing a doubly stochastic $v$, and it has also been
successfully applied to various types of problems
\cite{gol96}, although the method does have some weak
points as will be discussed below.
%

\subsubsection{Weak points with SoftAssign}
\label{wpSASec}
In the SoftAssign approach, the iterative Sinkhorn procedure for
normalizing $v$ is problematic at low $T$. This can be seen by
assuming one has reached a stage where the matrix is close to being
doubly stochastic:
\beq
	M_{ij} = (1 + \alpha_i) v_{ij} (1 + \beta_j),
\eeq
where $v$ is the desired doubly stochastic matrix, while $\alpha_i,
\beta_j$ are assumed small. To linear order in $\alpha, \beta$, one
step of row normalization corresponds to $\alpha \to - v \beta$ in
matrix notation, while one step of column normalization yields $\beta
\to - v^\top \alpha$. Together, this gives $\beta \to v^\top v \beta$.

Hence, the asymptotic rate of convergence is determined by the
eigenvalues of the positive-definite matrix $U = v^\top v$. At a high
temperature, $U$ will be close to a uniform matrix with $1/N$
everywhere, while at a low $T$, it will be close to the identity
matrix. Consider a simple Ansatz for $U$: $U_{ij} = (1 - a)
\delta_{ij} + a/N$, with $0\le a \le 1$. The limit $a\to 1$
corresponds to high $T$, while $a\to 0$ emulates low $T$. $U$ can also
be written as $U = (1-a){\bf 1} + aP$, where ${\bf 1}$ is the identity
matrix, and $P$ is the projection matrix onto the uniform vector.

$U$ has two distinct eigenvalues: a single unit eigenvalue with a
uniform eigenvector $(1,1,\dots,1)$, and an $(N-1)$-fold degenerate
value $1-a$ with eigenvectors in the transverse space. The unit
eigenvalue is to be expected, reflecting the irrelevance of shuffling
a global factor between $a$ and $b$.

What is worse is that as $T\to 0$ ($a\to0$), also the other
eigenvalue, $1-a$, tends towards unity.  This is not a special feature
of this particular Ansatz for $v$, but a generic phenomenon: As $v$
approaches a proper assignment matrix, $v^{\top}v$ approaches the
identity matrix. This means that the normalization procedure
inevitably runs into convergence problems in the limit of low $T$.

Another drawback as compared to PPP is that in SoftAssign, the
elements of the assignment matrix $v$ have to be updated in {\em
synchrony}, and there is no obvious simple way to update it, say, one
row at a time. As a result, stability is not guaranteed at low $T$
even for a good solution, unless the cost function $H$ is carefully
tuned.

\subsubsection{Speeding up the iterative normalization}
\label{normSec}

In order to improve convergence for the normalization procedure in
SoftAssign at low $T$, it appears important to initialize the
multiplicative row and column vectors $a$ and $b$ carefully, so as to
leave as little as possible for the slow iterative procedure to do.

Obviously, to avoid overflow on the computer upon implementation of
\eq{M}, the effective cost matrix will have to be modified with
suitable row and column additions, $c_{ij} \to c_{ij} + \alpha_i +
\beta_j$, to ensure that the smallest elements be zero and the rest
positive. This can be done, e.g., by first subtracting from the
elements in each row the lowest element in that row, and then
subtracting from the elements in each column the lowest element in
that column.

This measure does not suffice, however, to guarantee a proper starting
point. Consider $N=3$ and the cost matrix
$c=\matthree{1&0&0\\0&1&1\\0&1&1}$. At zero $T$ this yields
$M=\matthree{0&1&1\\1&0&0\\1&0&0}$, and the normalization procedure
will be caught in an eternal loop, alternating forever between the two
states $\matthree{0&1/2&1/2\\1&0&0\\1&0&0}$ and
$\matthree{0&1&1\\1/2&0&0\\1/2&0&0}$. In fact, this $M$ is not
normalizable with finite row and column factors!

Obviously, it is not enough to ensure at least one zero in each row and
each column of $c$. The zeros must be arranged such that at least one
combination of them will correspond to a one-to-one matching between
rows and columns. Finding such a modification with otherwise
non-negative elements corresponds precisely to solving the associated
(effective) linear assignment problem.

Thus, we suggest using e.g. the Hungarian method to transform $c$ to
the proper form, in order to guarantee a normalizable $M$ for $T\to0$.

Even this step will not guarantee a fast convergence. As a second
example, consider $N=2$ and an effective cost matrix $c =
\mattwo{0&0\\2&0}$, obviously in the proper form. The corresponding
$M$ at $T=0$ will be $\mattwo{1&1\\0&1}$. If this is handed to the
normalization procedure at a low $T$, the approach to the final doubly
stochastic matrix, $v = \mattwo{1&0\\0&1}$, will be merely harmonic.

The situation is considerably improved, by in addition carefully {\em
balancing} the cost matrix $c$, to ensure also a maximal number of
non-zero elements (while maintaining a sufficient set of zeros), with
the smallest of these being as large as possible. This can be done in
polynomial time, using a recursive procedure.
For the $N=2$ case above, the balancing step will yield
$c=\mattwo{0&1\\1&0}$, giving $M=\mattwo{1&0\\0&1}$ at zero $T$, and
convergence is immediate.

Still, there will be cases for $N>2$ where the procedure will be
caught in a slowly converging sequence -- this is inevitable, due to
the unit eigenvalues at zero $T$ --, but in this way the most obvious
traps are avoided.
%

An additional improvement is possible when using the Hungarian
algorithm for preprocessing the cost matrix, by exploiting that it
gives a preferred matching. This can be used to modify the
normalization process to improve the convergence rate by updating of
matched row-column pairs simultaneously. Especially in the low $T$
region, where the matched elements unambiguously define a selected
assignment, and the corresponding elements of $v$ will be close to
unity and the rest small, this noticeably speeds up the normalization.

Thus, the normalization constraints for a coupled row-column pair
$i,j$ of a modified $M$ read
\beqa 
\nonumber
a_i\left( \sum_{k \neq j} M_{ik} + M_{ij} b_j \right) &=& 1,
\\
\\
\nonumber
b_j \left( \sum_{k\neq i} M_{kj} + a_i M_{ij}\right) &=& 1,
\eeqa 
and are simultaneously satisfied by $a_i = x / A_i$
and $b_j = x / B_j$, where $A_i = \sum_{k\neq
j}M_{ik}$, $B_j = \sum_{k\neq i}M_{kj}$ and
\beq
	x = \frac{\sqrt{A_i B_j \left(4 M_{ij} + A_i B_j\right)}
	- A_i B_j} {2 M_{ij}}.
\eeq
Each matched row-column pair of $M$ in turn is updated in this manner
and the process is repeated until the result is close to being a
doubly stochastic matrix. We will refer to this normalization scheme
as {\em coupled normalization}.

\subsubsection{Ensuring a stable dynamics}
\label{SAsensibSec}

The second major problem with SoftAssign is the lack of a guarantee
for the stability of a solution in the low-$T$ limit;
this problem never appears e.g. in a properly handled system of Potts
MF spins with serial updating.

To remedy this, we will have to exploit the freedom of adding terms to
$H$ which are trivial on shell,
but nevertheless affect the MF dynamics.  One would at least like to
ensure that in the low-$T$ limit, SoftAssign turns into some form of
local optimization.

One possibility then is to demand that $H(v)$ be transformed into a
{\em concave} function of $v$ (in the subspace consistent with $v$
being doubly stochastic). This guarantees that the energy in
\eq{F_softass}
for $T\to 0$ will have a local minimum in the corner corresponding to
an optimal assignment.

A crude way to ensure concavity is to add a negative quadratic
diagonal term $-\left(\alpha/2\right) \sum_{ij} s_{ij}^2$ to $H$, with a large
enough coefficient $\alpha$. For a quadratic $H$ in particular,
concavity also ensures that the SoftAssign free energy
(\ref{F_softass}) becomes a Lyapunov function of the dynamics at a
fixed temperature \cite{ran97}. A disadvantage with this method is
that also non-solutions will be stabilized. Empirically, a smaller
diagonal addition often suffices to stabilize the dynamics.

A more advanced possibility is to employ problem-specific
modifications of $H$, adding suitable terms to ensure the stability of
a good solution.  It is therefore of interest to analyze what kinds of
generic additions are possible without altering the on-shell costs (or
merely adding a constant). We will refer to such terms as being {\bf
redundant}.

{\em Linear redundant terms}: Based on decomposing the defining
representation of the permutation group $P_N$, given by $s$, into the
direct sum of the trivial 1-dimensional irrep ${\bf e}$ and the fundamental
$(N-1)$-dimensional one ${\bf f}$ (see Appendix), the only possible linear
redundant additions are given by combinations of row or column sums of
$s$; in SoftAssign, such additions merely lead to a modification of
the initial row or column factors $a,b$ in \eq{v_aMb}, and have no
effect on the resulting doubly stochastic matrix $v$.

{\em Combinations of quadratic and linear terms}: The possibilities
here stem from the decomposition of the reducible representation of
$P_N$ defined by the symmetric direct product of the defining
representation ($\Leftrightarrow {\bf e}+{\bf f}$) with itself, as discussed in
the Appendix. The contributions stemming from the trivial direct
product of the ${\bf e}$ part with itself or with ${\bf f}$ give nothing useful,
corresponding to quadratic terms involving row or column sums of $s$,
completely equivalent to the corresponding linear terms with the row
or column sums replaced by 1.

The possibly interesting terms come from ${\bf f}\times{\bf f}={\bf e}+{\bf f}+{\bf s}+{\bf a}$. As
discussed in the Appendix, the symmetric part of this consists of a
part ${\bf a}$, antisymmetric in both row and column indices, yielding no
redundant terms, and a part containing ${\bf e}+{\bf f}+{\bf s}$, symmetric in both
row and column indices, which yields quadratic terms that vanish on
shell, or equal a constant, or a linear combination of the elements of
$s$.  In a slighly disguised form, they correspond to the on-shell
identities
\beqa
\nonumber
	s_{ij}^2 - s_{ij} &=& 0,
\\
	s_{ij} s_{il} &=& 0, \;\; \mbox{for } j \ne l,
\\
\nonumber
	s_{ij} s_{kj} &=& 0, \;\; \mbox{for } i \ne k.
\eeqa
Group theory certifies that these identities suffice to generate all
possibly useful redundant additions to $H(s)$, that are at most
quadratic in $s$.
Although such additions to $H$ are identically zero on shell ($v\to
s$), as additions to $H(v)$ they will alter the properties of the
dynamics in SoftAssign, in terms of a modification of the expression
for the effective cost matrix $c = \partial H(v)/\partial v$.

Thus, the addition of a term proportional to the square of an element
$v_{ij}$ minus the element itself, modifies only the corresponding
element of $c$, by an addition proportional to $2 v_{ij} - 1$.
Adding the product of two elements of $s$ in the same row $i$ but in
different columns $j,l$ affects the corresponding two elements of the
effective cost matrix, with an addition to $c_{ij}$ proportional to
$v_{il}$, and vice versa.
Analogously, adding to $H$ a term involving the product of two
elements in the same column $j$ but in different rows $i,k$ yields the
addition to $c_{ij}$ of a term proportional to $v_{kj}$ and vice
versa.

\subsubsection{TSP-specific modifications}
\label{tspModSec}

In certain quadratic problems, such as the {\em travelling salesman problem}
(TSP), where a set of $N$ {\em sites} is to be cyclically visited in an
optimal order, the cost function has the particular structure
\beq
	H(s) = \half \mbox{Tr} \left( s D s^{\top} X \right),
\label{tspH}
\eeq
with $D, X$ a pair of symmetric $N\times N$ matrices, vanishing on the
diagonal.

For TSP, $D$ is the pair-distance matrix, with $D_{ab}$ defining the
distance between sites $a,b$, while $X$ defines the cyclic tour
sequence neighborhood, $X_{ij} = \delta_{i,j+1} + \delta_{i,j-1}$,
such that $H$ measures the total {\em tour length}.

Concavity in the subspace consistent with $s$ being doubly stochastic,
of the direct product matrix $A = D \times X$, then corresponds to one
of $D$ or $X$ being positive-semidefinite, and the other
negative-semidefinite, each in the subspace orthogonal to $e =
(1,1,1\dots 1)$. This can be ensured by suitable diagonal additions to
$D$ and $X$ separately,
\beq
	D \to D + \alpha \one ,\; X \to X + \gamma \one,
\eeq
with $\alpha$ and $\gamma$ of opposite signs. The diagonal additions
to $D$ and $X$ implies adding terms to $H$ of the form discussed in
the previous subsection. All vanish on shell except for the $\alpha
\times \gamma$ term, which evaluates to a simple constant.

For the case of TSP, often $D$ is already negative-definite in the
transverse subspace, and a suitable addition to $X$ suffices. $X$ is
easily diagonalized by means of a discrete Fourier transform, with the
spectrum given by $\lambda_k = 2 \cos(2\pi k/N)$, for $k =
1,\dots,N-1$. Thus, $\gamma = 2$ is required to make the modified $X$
positive-semidefinite. In practice, however, $\gamma = 1$ will suffice
to stabilize the dynamics in most cases; in the low-$T$ limit this is
just enough to secure the stability of assignments locally optimal
with respect to local changes in the ordering of visited sites.

\section{Tests on Simple Applications}
 
In order to illustrate the ideas discussed in the previous section, we
will here test the various improvements to the SoftAssign algorithm,
as applied to a set of small single-assignment problems.

The effects of the improvements to the normalization algorithm at low
temperatures are illustrated using the linear assignment problem.
For TSP, the use of a problem-specific stabilizing term is compared to
employing a negative quadratic term $-\left(\alpha/2\right) \sum_{ij}
s_{ij}^2$.

The SoftAssign algorithm used is described in
Fig.~\ref{SAalgorithm}.\footnote{For a more thorough description of
SoftAssign in general, we refer to \cite{ran96}} All experiments have
been performed on an 800MHz PentiumIII computer running Linux.

\begin{figure}[htb]
\begin{center}
\fbox{\parbox{8.5cm}{
\begin{itemize}
\item Initiate the elements of $v$ to random values close to $1/N$,
  and $T$ to a high value.
\item Repeat the following (a sweep), until the $v$ matrix has
        {\em saturated} (i.e. become close to a ($0$,$1$)-matrix):
 \begin{itemize}
 \item Calculate the effective cost matrix by means of $c_{ij}=\partial
   H/\partial v_{ij}$, possibly modify it with suitable row/column
   additions, and let $M_{ij}=exp(-c_{ij}/T)$.
 \item Normalize $M$ with the proper row and column factors to yield a
   doubly stochastic matrix, defining an updated $v$.
 \item Decrease $T$ slightly (typically by a few percent).
 \end{itemize}
 \item Extract the resulting solution candidate.
\end{itemize}
}}
\end{center}
\caption{A SoftAssign algorithm}
\label{SAalgorithm}
\end{figure}
\subsection{Speeding up the iterative normalization}
As discussed in section \ref{wpSASec}, the Sinkhorn normalization of
$M$, \eq{sinkhorn}, runs into trouble when the temperature is low and
the corresponding $v$ is close to an on-shell assignment matrix.

To probe the efficiency of each normalization scheme, we use random
linear assignment, \eq{Hlinass}, where the costs $c_{ij}\in[0,1]$ are
uniform random numbers. We investigate the number of iteration steps
needed before row and column sums of the modified $M$ matrix are in
the range $1 \pm \Delta_{max}$ with $\Delta_{max}$ a small number, and
measure the time used by the normalization scheme. This is done at a
set of decreasing temperatures such that $v_{ij} \approx 1/N,\forall
i,j$ for the higher values of $T$, while $v$ is nearly on shell for
the lower $T$ values.

\begin{figure}[htb]
\begin{center}
\mbox{\includegraphics{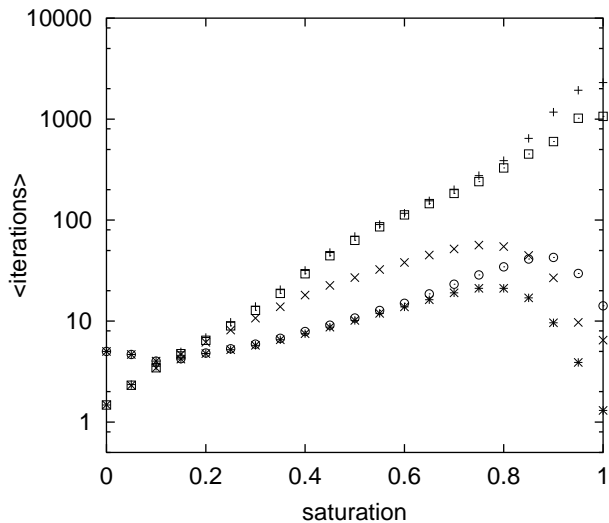}}
\end{center}
\caption{Number of normalization iterations used versus
saturation. The data consist of averages from runs on 100 random
linear assignment problems, binned into different values of the
saturation. The normalization procedure is continued until all row and
column sums are within $1\pm\Delta_{max}$ with $\Delta_{max}=0.01$.
The plot shows Plain Sinkhorn (+), Hungarian+Sinkhorn ($\Box$),
Hungarian+CoupledNorm ($\odot$),
Hungarian+Balancing+Sinkhorn ($\times$), and
Hungarian+Balancing+CoupledNorm (*).  }
\label{linAssItRes}
\end{figure}

\begin{figure}[htb]
\begin{center}
\mbox{\includegraphics{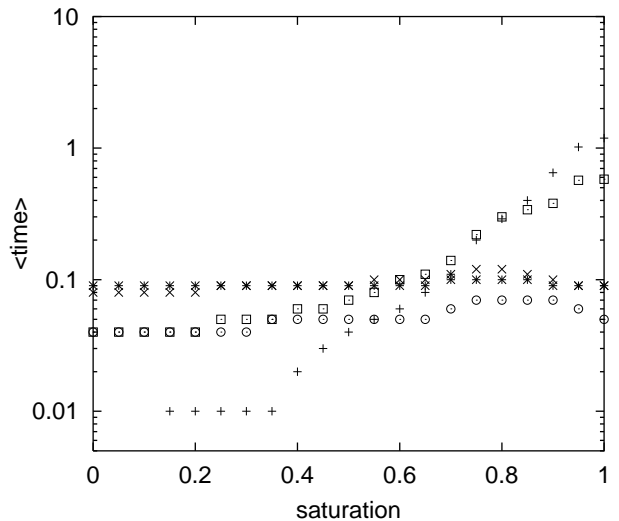}}
\end{center}
\caption{Time used for normalization, including preprocessing of the
cost matrix, versus saturation. The data consists of averages from
runs on 100 random linear assignment problems, binned into different
values of the saturation. The normalization procedure is continued
until all row and column sums are within $1\pm\Delta_{max}$ with
$\Delta_{max}=0.01$.
The plot shows Plain Sinkhorn (+), Hungarian+Sinkhorn ($\Box$),
Hungarian+CoupledNorm ($\odot$),
Hungarian+Balancing+Sinkhorn ($\times$), and
Hungarian+Balancing+CoupledNorm (*).  All times are measured in
seconds.}
\label{linAssTimeRes}
\end{figure}

We compare the following schemes described in section \ref{normSec}.
\begin{enumerate}
\item Plain Sinkhorn. Preprocess $c$ by first for each row subtracting
the smallest element, and then doing the same for each column. Then the
Sinkhorn row and column normalization (\eq{sinkhorn}) is applied on
the resulting $M$.
\item Hungarian+Sinkhorn. Preprocess $c$ using the Hungarian
algorithm. Then normalize $M$ using Sinkhorn.
\item Hungarian+Balancing+Sinkhorn. Preprocess $c$ using the
Hungarian and the balancing algorithms. Sinkhorn normalization of $M$.
\item Hungarian+CoupledNorm. Preprocess $c$ as in 2, then apply
the coupled row-column normalization described in section
\ref{normSec}.
\item Hungarian+Balancing+CoupledNorm. Same preprocessing of $c$ as in 3,
then coupled row-column normalization.
\end{enumerate}
Figures \ref{linAssItRes} and \ref{linAssTimeRes} show statistics from
100 linear assignment problems of size $N=100$. The data is binned for
different values of the {\em saturation}, $\Sigma =
\left(1/N\right)\sum_{ij}v_{ij}^2$, representing different temperature
regions. At high temperatures the saturation is close to $1/N$, while
it approaches one in the low temperature region.
The annealing is continued until the saturation becomes larger than
0.999, but is also aborted when the number of normalization steps
exceeds a maximal value of 20000 iteration steps for three consecutive
temperatures (which only happened for the plain Sinkhorn approach as
discussed below); these data points are not included when calculating
the averages in Figs. ~\ref{linAssItRes} and ~\ref{linAssTimeRes}.

The results illustrate the efficiency of the different normalization
methods used on $M$, and the effect of preprocessing of the cost
matrix $c$.
As can be seen in Fig.\ \ref{linAssItRes} the number of iterations
needed with plain Sinkhorn grows by several orders of magnitude in the
low temperature region, where the saturation approaches 1.
What is not revealed in the figure is that the plain Sinkhorn scheme
failed to normalize $M$ at low temperatures in {\em all} of the tested
problem instances -- because of the limited resolution on the
computer, small elements in $M$ become zero where the corresponding
elements in $v$ should be one. The Sinkhorn scheme then gets stuck in
an eternal loop, failing to produce a doubly stochastic $v$.
The failure occurred at high enough temperatures that the saturation
was below our limit of 0.999, and the algorithm had to be aborted as
described above.

This problem can be avoided by either interrupting the annealing at an
earlier stage, and extracting a solution from the unsaturated $v$
matrix, or by adding additional redundant terms to the Hamiltonian, at
the cost of a lower performance.
However, to guarantee that an arbitrary cost matrix yields a doubly
stochastic $v$, preprocessing of the cost matrix is essential.

Using a Hungarian preprocessor on the cost matrix ensures that the
initial $M$ has element values equal to one on a permutation, and
values between zero and one on the other elements. This guarantees
that at least the selected permutation survives in the normalizing
process, and a doubly stochastic matrix $v$ will always result. Though
this sounds appealing, our tests reveal that this does not
substantially decrease the number of Sinkhorn iterations as compared
to the plain Sinkhorn approach, as seen in Fig.~\ref{linAssItRes}.

To avoid the extreme increase of the number of normalization
iterations in the low temperature region one can apply balancing of
the cost matrix $c$, or use the coupled normalization approach, both
described in section \ref{normSec}. Applying either one (or both) of
the methods decreases the number of normalization iterations.  This is
evident in Fig.\ \ref{linAssItRes}, especially in the low temperature
region where the saturation approaches one.

Applying the Hungarian and balancing methods to the cost matrix $c$
leads to an increased time used to produce an initial $M$, which is
revealed in Fig.\ \ref{linAssTimeRes}. The total time for the
algorithm is dominated by the time spent on Hungarian and
balancing. This time is nevertheless far exceeded by the time required
with the plain Sinkhorn approach in the low temperature region.

The Hungarian method is known to scale in time with problem size as
$O(N^3)$ \cite{pap98}, and the balancing routine empirically appears
to behave similarly. This is to be compared to the time it takes to
calculate the effective cost matrix, which e.g. for TSP is also an
$O(N^3)$ procedure, while for generic quadratic assignment it scales
as $O(N^4)$.

\subsection{Stable dynamics in TSP}

One of the most studied combinatorial optimization problems is
TSP. Deterministic annealing has been applied to it using both PPP and
SoftAssign \cite{pet90,gol96} and we refer to these articles for
a more thorough description of the implementation on TSP. Here we will use 
TSP as an example where the choice of stabilizing term needed by SoftAssign
indeed influences the performance.
 
The standard assignment-matrix Hamiltonian for TSP is given in
equation \eq{tspH}. In addition to this an extra stabilizing term is
needed. We have compared the addition of a {\em generic stabilizer} in
the form of a diagonal quadratic term,
\beq
\label{diagQuad}
	H_A = -\frac{\alpha}{2}\sum_{ia}s_{ia}^2,
\eeq
as proposed in the literature \cite{gol96},
with a {\em problem-specific stabilizer} as discussed in section
\ref{tspModSec}($X \to X + \gamma \one$),
\beq
\label{diagX}
	H_B = \frac{\gamma}{2}\sum_{iab}s_{ia}s_{ib}D_{ab}.
\eeq

Throughout our experiments we have used the values 1.0 for both
$\alpha$ and $\gamma$. The $\alpha$ value is slightly smaller than the
value 1.4 used in \cite{gol96}. A $\gamma$ value of 1.0 ensures the
stability with respect to local changes of the ordering of visited
sites as discussed in section \ref{tspModSec}, but does not always
suffice to produce a proper assignment. When this happens (in about
1\% of the tested problems) the algorithm is restarted, initialized
with a new $v$. Typically, one restart is sufficient to find a proper
assignment matrix.
 
\begin{figure}[htb]
\begin{center}
\mbox{\includegraphics{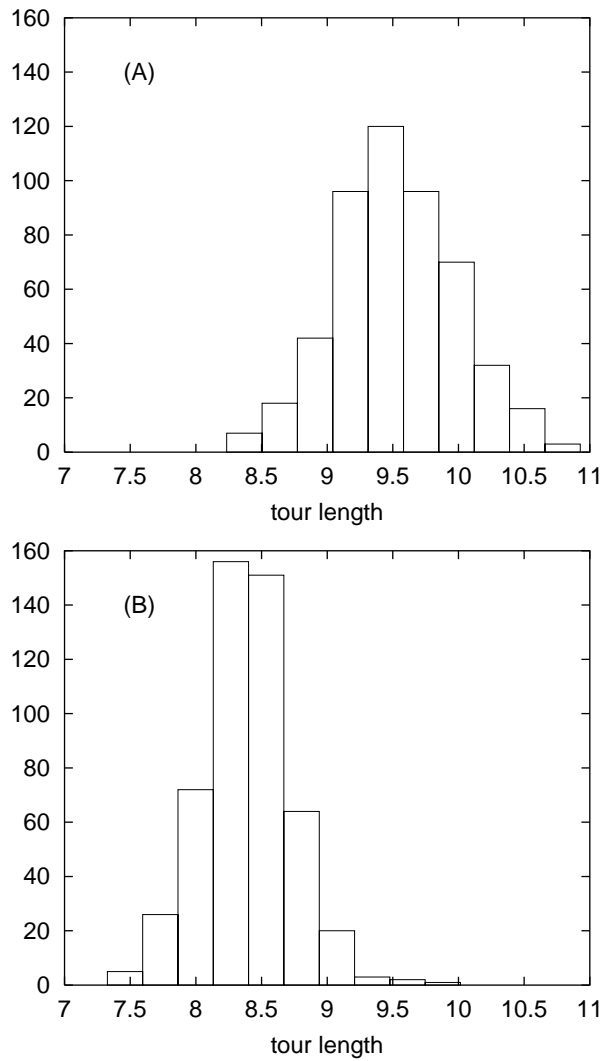}}
\end{center}
\caption{Tour lengths for 500 random two-dimensional Euclidean TSP
problems of size 100, using
({\bf A}) the generic stabilizer (\ref{diagQuad}) with $\alpha =
1.0$,
({\bf B}) the problem-specific stabilizer (\ref{diagX}) with
$\gamma = 1.0$.}
\label{tspRes}
\end{figure}

We studied random TSP problems where the sites were uniformly
generated in the two-dimensional unit square. In Fig.\ \ref{tspRes}
the tour lengths from 500 problems of size 100 are shown.

The generic stabilizer (\eq{diagQuad}) works by enhancing already
large spin elements. Due to this, $v$ saturates faster (towards an
assignment matrix) in the course of the annealing. Since this effect
is not as pronounced with the problem-specific stabilizer
(\eq{diagX}), equal annealing parameters will not lead to equal time
used by the algorithms.
Instead, parameters are chosen such that the respective performances
are not too far from optimal and the times used are comparable.  We
have used a slower annealing, $T \to T/1.01$, for the generic
stabilizer, and also allowed it to use up to 5 sweeps per temperature
if the maximal change in a spin components is larger than 0.01. An
even slower annealing would not lead to a considerably better
performance, in spite of the increase in time. With the
problem-specific stabilizer, we have used the annealing rate $T \to
T/1.05$, with one sweep per temperature.

With the generic stabilizer, the average tour length was 9.53 and the
average time used 1.53~s. With the problem-specific one, corresponding
values where 8.39 and 3.74~s (including possible restarts).  Thus,
performance-wise, the problem-specific stabilizer is superior, while
the increase in time used can be attributed to the slower saturation
-- this might be avoided by adding a small generic stabilizer term.

\section{Conclusions}

We have investigated the possibilities of defining a deterministic
annealing approach to nonlinear assignment problems, in anology to
existing algorithms for Ising and Potts systems.

We have analyzed a proper variational approach, where the problem cost
function is approximated by a variational cost, linear in the
assignment matrices. For a single assignment problem this allows for
an iterative scheme to minimize the variational free energy at a given
temperature. Combined with annealing, this can be used as a
deterministic annealing algorithm.

As an aside, the generalization to multiple assignment problems is
straightforward. Assuming additive linear contributions to the
variational cost from the different assignments leads to a
mean-field-like approximation with a factorized Boltzmann
distribution, and the variational parameters for the individual
assignments can be updated in a serial manner.

A major problem with the proper variational approach, however, is that
it requires the calculation of permanents, which needs exponential
time (in problem size). This implies that considering this as a
general heuristic for large nonlinear assignment problems is not
feasible.

Abandoning the quest for a proper variational method for nonlinear
assignment, we have also studied Potts-based methods as a more
promising alternative, although {\em per se} not tailored for
assignment problems. The currently most appealing method of this type
is the SoftAssign algorithm, and we have proposed some improvements to
it.

The Sinkhorn normalization procedure used in SoftAssign runs into
convergence problems at low temperatures. We present arguments why
this is unavoidable, and propose proper adjustments of the effective
cost matrix to reduce the effect. The application of a Hungarian
preprocessing to the effective cost matrix guarantees that the
Sinkhorn procedure always produces a doubly stochastic matrix. An
additional balancing of the cost matrix decreases the number of
iteration needed by the normalization. In addition we devise an
alternative normalization procedure which is easily implemented when a
Hungarian preprocessing is used. It is superior to the Sinkhorn
procedure at low temperatures.
We have experimentally confirmed these statements by implementation of
SoftAssign on random instances of linear assignment.
With other problem ensembles, however, we have experienced varying
effects of the improvements, in some cases they appear essential,
while in others they are more or less superfluous.

Another problem with the SoftAssign approach is the lack of guarantee
for stability in the low temperature region: Solutions may not be
stable. This problem can be resolved by adding to the Hamiltonian a
stabilizer -- a redundant term that affects the dynamics without
altering the on-shell cost. We have used arguments from group theory
to determine the possible types of redundant additions that are at
most quadratic in the spin components. As an example we discuss how
such redundant terms can be used for the travelling salesman problem,
and propose a TSP-specific stabilizing term different from the generic
one normally used with SoftAssign. In numerical experiments we show
that this enhances the performance.

\section{Acknowledgements}

This work was in part supported by the Swedish Foundation for
Strategic Research.



\appendix
\section{Basic Group Theory for $P_N$}
Here we will briefly review elements of the basic group theory for the
permutation group of $N$ elements, $G\equiv P_N$.

\subsection{Representations and irreps}

$G$ has a finite number of inequivalent irreducible representations,
or irreps; the squares of their dimensions sum up to the size $V$ of
the group, $V\equiv N!$. If $r$ labels an irrep, let $d_r$ be its
dimension, and let the associated $d_r\times d_r$ matrices be denoted
$u^r(g)$. We have $\sum_r d_r^2 = V$.

A $D$-dimensional reducible representation $\{U(g)\}$ can be
decomposed into the direct sum of (not necessarily distinct) irreps
$\{r_{\mu}\}$ as
\beq
\label{U_PPu}
	U_{ij}(g) = \sum_{\mu} \sum_{k,l} P^{\mu}_{ik} P^{\mu}_{jl} u^{r_{\mu}}_{kl}(g),
\eeq
or, in matrix form,
$U(g) = \sum_{\mu} P^{\mu} u^{r_{\mu}}(g) P^{\mu\top}$,
where $P^{\mu}$ is a ($g$-independent) $D\times d_{r_{\mu}}$ matrix
that projects out the part of a vector that belongs to the associated
irrep $r=r_{\mu}$. It can be seen as a submatrix of an orthogonal
$D\times D$ matrix $V$, used to similarity transform $U$ to an
explicitly blocked form $U_B$, $U(g) = V U_B(g) V^{\top}$.

The ortogonality of $V$ implies the following properties for the
matrices $P^{\mu}$:
\beqa
	&& \sum_{\mu} P^{\mu} P^{\mu\top} = {\bf 1}_D,
\\
	&& P^{\mu\top} P^{\nu} = \delta_{\mu\nu} {\bf 1}_{d_{r_{\mu}}},
\eeqa
where ${\bf 1}_d$ denotes the $d\times d$ identity matrix. Inverting
the similarity transform, \eq{U_PPu} is equivalent to
\beq
\label{PPU}
	\sum_{ij} P^{\mu}_{ik} P^{\nu}_{jl} U_{ij}(g) = \delta_{\mu\nu}
	u^{r_{\mu}}_{kl}(g),
\eeq
or, in matrix form,
$P^{\mu\top} U(g) P^{\nu} = \delta_{\mu\nu} u^{r_{\mu}}(g)$,
expressing the similarity transform of $U$ to blocked form, with
$\mu,\nu$ labeling respectively the row and column block.

The assignment matrices $s$ define a particular $N$-dimensional
representation (the defining representation) of $P_N$. It is {\em
reducible} if $N>1$, being the direct sum ${\bf e}+{\bf f}$ of two irreps, where
${\bf e}$ is the trivial one-dimensional irrep with $u^{{\bf e}}\equiv 1$, while
${\bf f}$ is a non-trivial $(N-1)$-dimensional irrep, the {\em fundamental}
representation. For the ${\bf e}$ part, e.g., the corresponding $N\times
1$-dimensional projection matrix is given by $P^{{\bf e}}_{ik}=1/\sqrt{N}$.

\subsection{Irrep Expansion}

Due to the identity $\sum_r d_r^2 = V$, there are as many distinct
matrix elements in the inequivalent irreps as there are elements in
the group. As is well known, these elements form a complete orthogonal
basis in group space, as expressed by
\beqa
	\sum_{g\in G} u^r_{ij}(g) u^s_{kl}(g) = \frac{V}{d_r}
	\delta_{r,s} \delta_{i,k} \delta_{j,l}
	&& \mbox{(orthogonality)},
\\
	\sum_r \sum_{i,j=1}^{d_r} d_r u^r_{ij}(g) u^r_{ij}(h) = V \delta_{g,h}
	&& \mbox{(completeness)}.
\eeqa
Thus, any function $F$ over the group can be expressed in a unique way
as a linear combination of the irrep elements,
\beq
\label{GT}
	F(g) = \sum_r \sum_{ij} C^r_{ij} u^r_{ij}(g),
\eeq
where $C$ are coefficients, and $u^r(g)$ is the orthogonal matrix
representing the group element $g$ in the irrep $r$.  Due to the
completeness, \eq{GT} can be inverted to yield the coefficients
uniquely as
\beq
\label{GTinv}
	C^r_{ij} = \frac{d_r}{N!} \sum_g u^r_{ij}(g) F(g).
\eeq
%

\subsubsection{Linear expressions in $s$}

\Eq{PPU} can be interpreted as follows: Independently of the group
element $g$, certain linear combinations of the elements of the matrix
$U(g)$ representing $g$ in a reducible representation $R$, will be
identical to the elements of the orthogonal matrix $u^r(g)$
corresponding to an irrep $r$ that appears in $R$; certain other
linear combinations (corresponding to $\mu\ne\nu$) will
vanish. Together, these span a complete basis in the space of all
possible linear combinations. This can be used to identify the
redundant linear or quadratic expressions in the assignment matrix
$s$.

A {\em linear} function of the assignment matrix $s$ can have
non-vanishing coefficients only for $r={\bf e}$ and $r={\bf f}$ in its expansion
(\ref{GT}).\footnote{Which shows that the most general cost function
is not linear in $s$.}
Separating the ${\bf e}$ and ${\bf f}$ parts of $s$ yields
\beq
	s_{ij} = 1/N + \sum_{kl} P^{{\bf f}}_{ik} u^{{\bf f}}_{kl} P^{{\bf f}}_{jl}.
\eeq
The different versions of \eq{PPU} then yield a set of identities,
\beqa
	\sum_{ij} s_{ij}(g) &=& N,
\\
	\sum_{ij} s_{ij}(g) P^{{\bf f}}_{jl} &=& 0,
\\
	\sum_{ij} P^{{\bf f}}_{ik} s_{ij}(g) &=& 0,
\\
	\sum_{ij} P^{{\bf f}}_{ik} s_{ij}(g) P^{{\bf f}}_{jl} &=& u^{{\bf f}}_{kl}(g).
\eeqa
The first three of these express in a slightly disguised form the
constraints of unit row and column sums in $s$.

\subsubsection{Quadratic expressions in $s$.}

A quadratic expression in $s$ means a linear combination of products
$U_{ik,jl} \equiv s_{ij} s_{kl}$, defining the direct
product representation $({\bf e}+{\bf f})\times({\bf e}+{\bf f})$, considering the pair of
row indices $\{i,k\}$ as a composite row index, and the pair of column
indices $\{j,l\}$ likewise as a composite column index.

It is reducible, and the corresponding version of \eq{PPU} reads
\beq
\label{PPU_ss}
	\sum_{ik,jl} Q^{\mu}_{ik,m} s_{ij}(g) s_{kl}(g) Q^{\nu}_{jl,n} =
	\delta_{\mu\nu} u^{r_{\mu}}_{mn}(g).
\eeq
The non-trivial part comes from ${\bf f}\times{\bf f}$, which reduces to
${\bf e}+{\bf f}+{\bf s}+{\bf a}$ (for $N>2$), where ${\bf s}$ and ${\bf a}$ are two new irreps, of
dimensionalities $d_{{\bf s}}=N(N-3)/2$ and $d_{{\bf a}}=(N-1)(N-2)/2$.

Due to the obvious symmetry of $U$, $U_{ik,jl} \equiv U_{ki,lj}$, it
is natural to divide the resulting irreps in two sets, according to the
symmetry with respect to swapping the index pair $ik$ in
$Q^{\mu}_{ik,m}$.
Thus, the {\em symmetric} part of ${\bf f}\times{\bf f}$ contains ${\bf e}+{\bf f}+{\bf s}$,
while the {\em antisymmetric} part contains ${\bf a}$ alone.

With opposite symmetry type in the row and column parts, the LHS of
\eq{PPU_ss} vanishes identically. Thus, the interesting parts require
$\mu,\nu$ to correspond to the same type of symmetry.  The
antisymmetric part contains only the nontrivial part $\mu=\nu={\bf a}$,
yielding $u^{{\bf a}}$. In the symmetric part, the $\mu=\nu={\bf s}$ part
similarly yields $u^{{\bf s}}$, while the remaining combinations yield
products of elements of $s$ for which the RHS will either vanish
($\mu\ne\nu$), or yield a constant ($\mu=\nu={\bf e}$), or a linear
combination of the elements of $s$ ($\mu=\nu={\bf f}$); the LHS then
defines candidates for redundant quadratic additions to a cost
function.

\end{document}